\documentstyle[12pt,epsfig,graphics,cite,amssymb,amsmath,bm,cool,caption]{article}
\begin{document}\bibliographystyle{plain}\begin{titlepage}
\renewcommand{\thefootnote}{\fnsymbol{footnote}}\hfill
\begin{tabular}{l}HEPHY-PUB 941/14\\UWThPh-2014-23\\October
2014\end{tabular}\\[2cm]\Large\begin{center}{\bf THE SPINLESS
RELATIVISTIC YUKAWA PROBLEM}\\[1cm]\large{\bf Wolfgang
LUCHA\footnote[1]{\normalsize\ \emph{E-mail address\/}:
wolfgang.lucha@oeaw.ac.at}}\\[.3cm]\normalsize Institute for High
Energy Physics,\\Austrian Academy of Sciences,\\Nikolsdorfergasse
18, A-1050 Vienna, Austria\\[1cm]\large{\bf Franz
F.~SCH\"OBERL\footnote[2]{\normalsize\ \emph{E-mail address\/}:
franz.schoeberl@univie.ac.at}}\\[.3cm]\normalsize Faculty of
Physics, University of Vienna,\\Boltzmanngasse 5, A-1090 Vienna,
Austria\\[2cm]{\normalsize\bf Abstract}\end{center}\normalsize

\noindent Noticing renewed or increasing interest in the
possibility to describe semirelativistic bound states (of either
spin-zero constituents or, upon confining oneself to spin-averaged
features, constituents with nonzero spin) by means of the spinless
Salpeter equation generalizing the Schr\"odinger equation towards
incorporation of effects caused by relativistic kinematics, we
revisit this problem for interactions between bound-state
constituents of Yukawa shape, by recalling and applying several
well-known tools enabling to constrain the resulting~spectra.
\vspace{3ex}

\noindent\emph{PACS numbers\/}: 03.65.Pm, 03.65.Ge, 12.39.Pn,
11.10.St\vspace{2ex}

\noindent\emph{Keywords\/}: relativistic bound states,
Bethe--Salpeter formalism, spinless Salpeter equation,
Rayleigh--Ritz variational technique, Yukawa potential

\renewcommand{\thefootnote}{\arabic{footnote}}\end{titlepage}

\section{Introduction: Describing Relativistic Bound States}The
\emph{spinless Salpeter equation\/}, frequently employed for a
quantum-theoretic description of semirelativistic bound states, is
the eigenvalue equation of a Hamiltonian $H$ combining the
relativistic free energy $T$ of the bound-state constituents with
some interaction potential $V.$ It is encountered as penultimate
step (the ultimate step being the Schr\"odinger equation) in the
course of a nonrelativistic reduction of the homogeneous
Bethe--Salpeter equation \cite{BSEa,BSEb,BSEc} as the
Lorentz-covariant quantum-field-theoretic approach to bound
states. For a system~of two particles of equal masses $m,$
relative momentum $\bm{p},$ and relative coordinate $\bm{x},$ $H$
reads\footnote{To simplify notation, we use units natural in
relativistic quantum theory or particle physics:~$\hbar=c=1.$}
\begin{equation}H\equiv T(\bm{p})+V(\bm{x})\ ,\qquad T(\bm{p})
\equiv2\,\sqrt{\bm{p}^2+m^2}\ .\label{Eq:H}\end{equation}The
hardly avoidable nonlocality of $H$ inhibits to find the exact
eigensolutions analytically.

In this study, we revisit what we call the spinless relativistic
Yukawa problem, posed by allowing $V$ to be the Yukawa potential,
a short-range, spherically symmetric potential~(i.e., $V=V(r),$
$r\equiv|\bm{x}|$) with coupling strength $y$ and shape determined
by a range parameter~$b$:\begin{equation}V(\bm{x})=V_{\rm
Y}(r)\equiv-y\,\frac{\exp(-b\,r)}{r}\ ,\qquad b>0\ ,\qquad y\ge0\
.\label{Eq:VY}\end{equation}Owing to the fact that it may be
understood to arise from the exchange of a single mediator of mass
$b$ between interacting particles, this potential is of paramount
importance for many subareas of physics. Like its limit for
$b\to0,$ the Coulomb potential, $V_{\rm Y}$ is singular~at $r=0.$
Our goal is to derive \emph{rigorous\/} constraints on the
resulting eigensolutions by standard tools.

\section{Energy Bounds for the Spinless Salpeter Equation}
\label{Sec:EB}Semiboundedness of a linear operator may be
established, possibly, by comparison. For any semibounded
operator, the coarse location of its eigenvalues can be restricted
variationally. Both lower and upper spectral limits may by found
by a tool dubbed envelope theory \cite{Lucha00-HO,Lucha01-DMAIa,
Lucha01-DMAIb,Lucha02-sum,Lucha02-DMAII}.

\subsection{Various Upper Bounds}For bound states described by the
spinless Salpeter equation, a couple of results of different
origin for upper limits to energy levels have been derived: some
of rather trivial nature,~e.g., those emerging from the
corresponding Schr\"odinger problem; or those subsumed under the
notion of (appropriately constructed) envelope theory; or those
arising from the variational characterization of eigenvalues of
operators provided by the minimum--maximum theorem.

\subsubsection{Nonrelativistic Kinematics: Straightforward
Schr\"odinger Upper Bound}\label{Sec:SUB}Both the concavity of the
square root in the relativistic kinetic energy $T(\bm{p})$
considered as a function of $\bm{p}^2$ and, for self-adjoint
$T(\bm{p}),$ the operator inequality \cite{Martin88}
$\left[T(\bm{p})-2\,m\right]^2\ge0$~yield
$$T(\bm{p})\le2\,m+\frac{\bm{p}^2}{m}\ .$$So, our
spinless-Salpeter Hamiltonian $H$ and its nonrelativistic limit,
$H_{\rm NR},$ must satisfy~the operator relation $H\le H_{\rm NR}$
and their eigenvalues $E_k$ and $E_{k,{\rm NR}},$ respectively,
the inequality $E_k\le E_{k,{\rm NR}}$ for all $k=0,1,2,\dots.$
This entails that the total number $N$ of spinless-Salpeter bound
states will not be less than the number $N_{\rm NR}$ of
Schr\"odinger bound states: $N\ge N_{\rm NR}.$

\subsubsection{Relativistic Kinematics: Variational Upper Limits by
Rayleigh and Ritz}\label{Sec:RUB}As consequence of the
minimum--maximum theorem \cite{MMT1,MMT2,MMT3}, the Rayleigh--Ritz
variational tool takes advantage of the fact that the lowest-lying
$d$ eigenvalues $E_k,$ $k=0,1,2,\dots,d-1,$ ordered according to
$E_0\le E_1\le E_2\le\cdots$ of a self-adjoint operator $H$
bounded from below (if still all situated below the onset of the
essential spectrum of $H$) are bounded from above by the $d$
similarly ordered eigenvalues $\widehat E_k,$ $k=0,1,\dots,d-1,$
of the operator $H$ restricted to a $d$-dimensional trial subspace
of the domain of $H$: $E_k\le\widehat E_k$ for all
$k=0,1,\dots,d-1,$ which allows us to localize eigenvalues by
upper bounds of increasing~tightness
\cite{Lucha:Q&Aa,Lucha:Q&Ab}~for~rising~$d.$

We find and have always found \cite{Lucha92:MEP,Lucha94:Como,
Lucha96:CCC,Lucha97:LagBd,Lucha99:talk1,Lucha99:talk2,Lucha04:TWR,
Lucha14:SRWSP,Lucha14:QCD@Work, Lucha14:SRHP} it advantageous to
span these finite-dimensional variational trial subspaces by a
basis the representations of which are known analytically in both
configuration and momentum space (related, of course, by Fourier
transformation): in this case, the expectation values of $H$ may
be analytically given by evaluating those of $T(\bm{p})$ in
momentum space and those of $V(\bm{x})$ in configuration space. By
spherical symmetry, each basis vector factorizes into the product
of a radial part and a spherical harmonic ${\cal Y}_{\ell
m}(\Omega)$ for angular momentum $\ell$ and projection $m$
depending on the solid angle $\Omega.$ Our (orthonormal)
configuration-space basis vectors use generalized-Laguerre
orthogonal polynomials \cite{AS,Bateman}:
\begin{align*}&\phi_{k,\ell
m}(\bm{x})=\sqrt{\frac{(2\,\mu)^{2\ell+2\beta+1}\,k!}
{\Gamma(2\,\ell+2\,\beta+k+1)}}\,|\bm{x}|^{\ell+\beta-1}
\exp(-\mu\,|\bm{x}|)\,L_k^{(2\ell+2\beta)}(2\,\mu\,|\bm{x}|)\,
{\cal Y}_{\ell m}(\Omega_{\bm{x}})\ ,\\
&L_k^{(\gamma)}(x)\equiv\sum_{t=0}^k\binom{k+\gamma}{k-t}
\frac{(-x)^t}{t!}\ ,\qquad k=0,1,2,\dots\ ,\qquad\mu\in(0,\infty)\
,\qquad\beta\in\!\left(-\frac{1}{2},\infty\right).\end{align*}Our
momentum-space basis vectors involve the hypergeometric function,
$F(u,v;w;z)$~\cite{AS}:\begin{align*}\widetilde\phi_{k,\ell
m}(\bm{p})&=\sqrt{\frac{(2\,\mu)^{2\ell+2\beta+1}\,k!}
{\Gamma(2\,\ell+2\,\beta+k+1)}}\,\frac{(-{\rm
i})^\ell\,|\bm{p}|^\ell}{2^{\ell+1/2}\,
\Gamma\!\left(\ell+\frac{3}{2}\right)}\\[1ex]&\times
\sum_{t=0}^k\,\frac{(-1)^t}{t!}\binom{k+2\,\ell+2\,\beta}{k-t}
\frac{\Gamma(2\,\ell+\beta+t+2)\,(2\,\mu)^t}
{(\bm{p}^2+\mu^2)^{(2\ell+\beta+t+2)/2}}\\[1ex]&\times
F\!\left(\frac{2\,\ell+\beta+t+2}{2},-\frac{\beta+t}{2};
\ell+\frac{3}{2};\frac{\bm{p}^2}{\bm{p}^2+\mu^2}\right){\cal
Y}_{\ell m}(\Omega_{\bm{p}})\ ,\\&F(u,v;w;z)\equiv
\frac{\Gamma(w)}{\Gamma(u)\,\Gamma(v)}\,\sum_{n=0}^\infty\,
\frac{\Gamma(u+n)\,\Gamma(v+n)}{\Gamma(w+n)}\,\frac{z^n}{n!}\
.\end{align*}

\subsection{Coulomb Lower Bounds to the Relativistic Yukawa
Problem}\label{Eq:LEB}Since $\exp(-b\,r)\le1,$ the potential
(\ref{Eq:VY}) is bounded from below by the Coulomb
potential~$V_{\rm C}$:$$V_{\rm C}(r)\equiv -\frac{\kappa}{r}\le
-y\,\frac{\exp(-b\,r)}{r}\equiv V_{\rm Y}(r)\qquad\mbox{for}\qquad
y\le\kappa\ .$$For the relativistic Coulomb problem, various
rigorous lower limits have been given \cite{Herbst,Martin89}:
\begin{itemize}\item Disregarding domain questions, the Hamiltonian
(\ref{Eq:H}) with Coulomb potential $V_{\rm C}(r)$ is essentially
self-adjoint for $\kappa\le1,$ with Friedrichs extension up to the
critical~coupling$$\kappa_{\rm c}=\frac{4}{\pi}=1.273239\dots\ ,$$
and, if $\kappa<\kappa_{\rm c},$ the spectrum of the Hamiltonian
$H,$ $\sigma(H),$ is bounded from below \cite{Herbst}:
\begin{equation}\sigma(H)\ge2\,m\,
\sqrt{1-\left(\frac{\kappa}{\kappa_{\rm c}}\right)^{\!2}}=2\,m\,
\sqrt{1-\left(\frac{\pi\,\kappa}{4}\right)^{\!2}}\ .\label{CLBH}
\end{equation}\item For a restricted range of coupling constants,
the spectral bound can be improved~\cite{Martin89}:
\begin{equation}\sigma(H)\ge2\,m\,
\sqrt{\frac{1+\sqrt{1-\kappa^2}}{2}}\qquad\mbox{for}\qquad\kappa
\le1\ .\label{CLBMR}\end{equation}\end{itemize}Clearly, even on
dimensional grounds the bounds (\ref{CLBH}) and (\ref{CLBMR}) have
to scale with the mass $m.$ So, (at least) for $y\le\kappa_{\rm
c},$ the operator (\ref{Eq:H}) with Yukawa potential (\ref{Eq:VY})
is bounded from~below. Hence, the minimum--maximum theorem applies
and using variational methods is justified.

\section{Relativistic Yukawa Problem: General Constraints}
\label{Sec:YP}Further, more Yukawa-specific issues worth
consideration are the allowed number of bound states (well-defined
in the Schr\"odinger case) or the existence of a critical coupling
constant.

\subsection{Rigorous Limit on the Number of Schr\"odinger Bound
States}\label{Sec:NBS}The nonrelativistic Yukawa problem can
accommodate only a finite total number of bound states. A rather
simple upper limit to this number $N_{\rm NR}$ has been derived by
Bargmann~\cite{Bargmann}:$$N_{\rm NR}\lneqq\frac{I\,(I+1)}{2}\
,\qquad I\equiv m\int\limits_0^\infty{\rm d}r\,r\,|V_{\rm
Y}(r)|=\frac{m\,y}{b}\ .$$

\subsection{\boldmath Boundedness from Below: Constraint on
Coupling Constant}\label{Sec:CCCC}The singularity at the origin
($r=0$) of the Yukawa potential $V_{\rm Y}(r)$ resembles the
one~of~the Coulomb potential $V_{\rm C}(r).$ Therefore, also for
the relativistic Yukawa problem the existence of a ground state
(or, in other words, of a finite eigenvalue $E_0>-\infty$ of the
Hamiltonian~$H$) requires the overall coupling constant in this
potential to be bounded from above. Since,~by Rayleigh's
principle, which is nothing but the minimum--maximum principle for
trial-space dimension $d=1,$ any expectation value $\langle
H\rangle$ of $H$ forms an upper limit to the bottom of the
spectrum of $H,$ demanding $\langle H\rangle$ to be finite,
$\langle H\rangle>-\infty,$ restricts the allowable couplings $y.$
An analytic limit is found by use of the simplest of our basis
states~($k=\ell=m=0,$~$\beta=1$),\begin{equation}
\phi_{0,00}(\bm{x})=\sqrt{\frac{\mu^3}{\pi}}\exp(-\mu\,|\bm{x}|)\
,\qquad\widetilde\phi_{0,00}(\bm{p})=\frac{\sqrt{8\,\mu^5}}{\pi}\,
\frac{1}{(\bm{p}^2+\mu^2)^2}\ ,\label{Eq:0001}\end{equation}
yielding, for the expectation values of kinetic energy $T(\bm{p})$
\cite{Lucha14:SRHP} and Yukawa potential~$V_{\rm
Y}(r),$\begin{align*}\langle
T(\bm{p})\rangle&=\frac{4}{3\,\pi\,(m^2-\mu^2)^{5/2}}\\
&\times\left[\mu\,\sqrt{m^2-\mu^2}\,(3\,m^4-4\,m^2\,\mu^2+4\,\mu^4)
+3\,m^4\,(m^2-2\,\mu^2)\,\ArcSec{\frac{m}{\mu}}\right],\\\langle
V_{\rm Y}(r)\rangle&=-\frac{4\,y\,\mu^3}{(b+2\,\mu)^2}\ .
\end{align*}We must inspect the limit $\mu\to\infty,$ where our
trial state (\ref{Eq:0001}) gets concentrated to the~spatial
origin, $r=0.$ In this case, $\langle H\rangle=\langle
T(\bm{p})\rangle+\langle V_{\rm Y}(r)\rangle$ becomes, by its
expansion about $\mu=\infty,$ $$\langle
H\rangle=\left(\frac{16}{3\,\pi}-y\right)\mu+y\,b+\frac{1}{\mu}
\left(\frac{16\,m^2}{3\,\pi}-\frac{3\,y\,b^2}{4}\right)
+O\!\left(\frac{1}{\mu^2}\right)\xrightarrow[\mu\to\infty]{}
-\infty\qquad\mbox{for}\qquad y>\frac{16}{3\,\pi}\ .$$Any coupling
$y$ large enough that the coefficient in the $O(\mu)$ term is
negative will inevitably cause collapse. To avoid this, we impose
the constraint (improvable by larger dimensions~$d$)
$$y\le\frac{16}{3\,\pi}=1.69765\dots\ .$$This constraint does not
depend on the Yukawa range $b.$ It thus holds without modification
also for the limit $b\to0$ of the Yukawa potential, the Coulomb
potential. There, of course, it cannot be as tight as the optimum
constraint $\kappa<\kappa_{\rm c}$ by Herbst
(cf.~Ref.~\cite[Theorem~2.1]{Herbst}).

\section{Application: Tentative Approximate Eigensolution}
\label{Sec:App2}Upon specifying the numerical values of the
parameters of the theory, the various concepts, findings, and
techniques introduced in Secs.~\ref{Sec:EB} and \ref{Sec:YP} may
now be imposed on the relativistic Yukawa problem. We are aware of
only two recently published investigations of the spinless
Salpeter equation with Yukawa potential
\cite{Hamzavi_IJMPE,Hassanabadi_CPB}, both of them aiming at
finding (of course, just approximate) analytical solutions by
replacing this bound-state equation by~a~``pseudo spinless
Salpeter equation,'' a --- by standard methods solvable
--- Schr\"odinger-like implicit eigenvalue equation, fabricated
by expansion of the relativistic free energy $T(\bm{p})$ up to
order $\bm{p}^4/m^4$ and a perturbative treatment of the
$O(\bm{p}^4)$ term spoiling the semiboundedness of~$H.$

For ease of comparison, we prefer to adopt the tools prepared
beforehand within~a~setup that has been discussed already in
earlier investigations. Among the two studies mentioned above,
Ref.~\cite{Hamzavi_IJMPE} presents more details. So, let's employ
also here the set of parameter~values used in
Ref.~\cite{Hamzavi_IJMPE}, that is, $m=5\;\mbox{fm}^{-1}$ for the
common mass of the bound-state constituents, $y=1$ for the Yukawa
coupling constant, and three different choices for the Yukawa
range $b.$ For the envisaged comparison of outcomes, we opt for
the largest among these three values, $b=0.01\;\mbox{fm}^{-1}$:
this choice clearly entails the largest deviation of the Yukawa
potential from the Coulomb case realized for $b=0.$ The value
$y=1$ respects the bound~derived~in~Sec.~\ref{Sec:CCCC}:
$$y=1<\frac{16}{3\,\pi}=1.69765\dots\ .$$Even the value
$b=0.01\;\mbox{fm}^{-1}$ is, however, so small that the Yukawa
problem under study is still close enough to its Coulomb limiting
case that the upper limit on the number~of bound states of the
nonrelativistic Yukawa problem given in Sec.~\ref{Sec:NBS} is
fairly large: $N_{\rm NR}<125250.$

The Coulomb lower limits on the relativistic Yukawa spectrum
require $\kappa\ge y$ (Sec.~\ref{Eq:LEB}). The smallest allowed
Coulomb coupling $\kappa=y=1$ then constrains the lowest
eigenvalue~by
$$\frac{E_0}{m}\ge\frac{\sqrt{16-\pi^2}}{2}=1.23798\dots\quad
\mbox{(Ref.~\cite{Herbst})}\ ,\qquad\frac{E_0}{m}\ge\sqrt{2}
=1.41421\dots\quad\mbox{(Ref.~\cite{Martin89})}\ .$$Letting
$m=5\;\mbox{fm}^{-1}$ thus implies, as lower bounds to the
ground-state energy eigenvalue $E_0$ or the associated binding
energy, $B_0\equiv E_0-2\,m,$ respectively, from Eq.~(\ref{CLBH}),
$E_0\ge6.19\;\mbox{fm}^{-1}$ and $B_0\ge-3.81\;\mbox{fm}^{-1}$
and, by the improvement (\ref{CLBMR}),
$E_0\ge7.07\;\mbox{fm}^{-1}$ and $B_0\ge-2.93\;\mbox{fm}^{-1}.$

For the \emph{lowest\/} bound states (labelled by the number of
zeros of the radial wave~function $n_r$ and the angular momentum
$\ell$) of the relativistic Yukawa problem, both upper bounds~of
Sec.~\ref{Sec:EB} on the \emph{binding\/} energies $B_k\equiv
E_k-2\,m,$ $k=0,1,2,\dots,d-1,$ are presented in
Table~\ref{Tab:RYP}.

\begin{table}[ht]\caption{Upper limits on the binding energies of
the (actually) lowest bound-state solutions of the spinless
Salpeter equation with Yukawa potential, for the parameters of
Ref.~\cite{Hamzavi_IJMPE}: the Laguerre bounds $\overline{B}$ of
Sec.~\ref{Sec:RUB} and the Schr\"odinger bounds $\overline{B}_{\rm
NR}$ of Sec.~\ref{Sec:SUB}. Each bound state is identified by both
its radial quantum number $n_r$ and its orbital~angular~momentum
quantum number $\ell$. Just as a starting point, we keep the
dimension $d$ of the variational trial space $D_d$ and both
variational parameters $\mu$ and $\beta$ fixed to the values
$d=25,$~$\mu=1,$~$\beta=1.$}\label{Tab:RYP}
\begin{center}\begin{tabular}{ccll}\hline\hline\\[-1.5ex]
\multicolumn{2}{c}{Bound state}&\multicolumn{1}{c}{Spinless
Salpeter equation}&\multicolumn{1}{c}{Schr\"odinger equation}
\\[1ex]$n_r$&$\ell$&\multicolumn{1}{c}{$\overline{B}(n_r,\ell)
\left[\mbox{fm}^{-1}\right]$}&\multicolumn{1}{c}{$\overline{B}_{\rm
NR}(n_r,\ell)\left[\mbox{fm}^{-1}\right]$}\\[1.5ex]\hline\\[-1.49ex]
0&0&$\qquad\quad-1.62087$&$\qquad-1.24002$\\
0&1&$\qquad\quad-0.31432$&$\qquad-0.30259$\\
0&2&$\qquad\quad-0.13085$&$\qquad-0.12909$\\
0&3&$\qquad\quad-0.06895$&$\qquad-0.06847$\\
0&4&$\qquad\quad-0.04070$&$\qquad-0.04052$\\
0&5&$\qquad\quad-0.02555$&$\qquad-0.02546$\\
0&6&$\qquad\quad-0.01683$&$\qquad-0.01649$\\[.39ex]
1&0&$\qquad\quad-0.37016$&$\qquad-0.30262$\\
1&1&$\qquad\quad-0.13413$&$\qquad-0.12913$\\
1&2&$\qquad\quad-0.06958$&$\qquad-0.06853$\\
1&3&$\qquad\quad-0.04094$&$\qquad-0.04060$\\
1&4&$\qquad\quad-0.02570$&$\qquad-0.02555$\\
1&5&$\qquad\quad-0.01669$&$\qquad-0.01659$\\
1&6&$\qquad\quad-0.01127$&$\qquad-0.01089$\\[.39ex]
2&0&$\qquad\quad-0.15029$&$\qquad-0.12915$\\
2&1&$\qquad\quad-0.07098$&$\qquad-0.06857$\\
2&2&$\qquad\quad-0.04128$&$\qquad-0.04065$\\
2&3&$\qquad\quad-0.02585$&$\qquad-0.02562$\\
2&4&$\qquad\quad-0.01680$&$\qquad-0.01667$\\
2&5&$\qquad\quad-0.01095$&$\qquad-0.01099$\\
2&6&$\qquad\quad-0.00685$&$\qquad-0.00721$\\[.39ex]
3&0&$\qquad\quad-0.07764$&$\qquad-0.06858$\\
3&1&$\qquad\quad-0.04201$&$\qquad-0.04069$\\
3&2&$\qquad\quad-0.02606$&$\qquad-0.02567$\\
3&3&$\qquad\quad-0.01690$&$\qquad-0.01674$\\
3&4&$\qquad\quad-0.01098$&$\qquad-0.01107$\\
3&5&$\qquad\quad-0.00734$&$\qquad-0.00730$\\
3&6&$\qquad\quad-0.00212$&$\qquad-0.00471$\\[.39ex]
4&0&$\qquad\quad-0.04535$&$\qquad-0.04071$\\
4&1&$\qquad\quad-0.02649$&$\qquad-0.02570$\\
4&2&$\qquad\quad-0.01704$&$\qquad-0.01679$\\
4&3&$\qquad\quad-0.01097$&$\qquad-0.01113$\\
4&4&$\qquad\quad-0.00646$&$\qquad-0.00737$\\
4&5&$\qquad\quad-0.00134$&$\qquad-0.00479$\\[1.2ex]
\hline\hline\end{tabular}\end{center}\end{table}\clearpage

\begin{table}[ht]\caption*{Table~\ref{Tab:RYP}, continued.}
\begin{center}\begin{tabular}{ccll}\hline\hline\\[-1.5ex]
\multicolumn{2}{c}{Bound state}&\multicolumn{1}{c}{Spinless
Salpeter equation}&\multicolumn{1}{c}{Schr\"odinger equation}
\\[1ex]$n_r$&$\ell$&\multicolumn{1}{c}{$\overline{B}(n_r,\ell)
\left[\mbox{fm}^{-1}\right]$}&\multicolumn{1}{c}{$\overline{B}_{\rm
NR}(n_r,\ell)\left[\mbox{fm}^{-1}\right]$}\\[1.5ex]\hline\\[-1.49ex]
5&0&$\qquad\quad-0.02839$&$\qquad-0.02572$\\
5&1&$\qquad\quad-0.01730$&$\qquad-0.01682$\\
5&2&$\qquad\quad-0.01094$&$\qquad-0.01117$\\
5&3&$\qquad\quad-0.00531$&$\qquad-0.00743$\\[.39ex]
6&0&$\qquad\quad-0.01848$&$\qquad-0.01684$\\
6&1&$\qquad\quad-0.01087$&$\qquad-0.01121$\\
6&2&$\qquad\quad-0.00438$&$\qquad-0.00747$\\[.39ex]
7&0&$\qquad\quad-0.01151$&$\qquad-0.01122$\\
7&1&$\qquad\quad-0.00349$&$\qquad-0.00750$\\[.39ex]
8&0&$\qquad\quad-0.00363$&$\qquad-0.00752$\\[1.2ex]
\hline\hline\end{tabular}\end{center}\end{table}

A brief comparison of the upper energy bounds listed in Table
\ref{Tab:RYP} with the corresponding numerical results presented
in Tables 1 and 2 of Ref.~\cite{Hamzavi_IJMPE} reveals a few
unexpected features, which clearly affects one's assessment of the
reliability of the approach followed in Ref.~\cite{Hamzavi_IJMPE}:
\begin{itemize}\item For pretty unclear reasons, the \emph{ground
state\/} of the pseudo spinless Salpeter equation, studied in
Ref.~\cite{Hamzavi_IJMPE}, characterized by vanishing radial and
orbital angular momentum quantum numbers, i.e., $n_r=\ell=0,$ is
missing in both Table 1 and Table 2 of Ref.~\cite{Hamzavi_IJMPE}.
As a matter of fact, in each sector of given angular momentum
$\ell$ the (nodeless) bound system with radial quantum number
$n_r=0$ didn't make it into the study~of
Ref.~\cite{Hamzavi_IJMPE}. Hence, for the sake of completeness,
comparability and compatibility with~our results for the upper
energy limits, Table \ref{Tab:AGS} discloses the binding energies
of all $n_r=0$ bound states accommodated by the pseudo
spinless-Salpeter equation in Eq.~(4) of
Ref.~\cite{Hamzavi_IJMPE}.

\begin{table}[hb]\caption{Binding energies $B_{\rm
p}(n_r=0,\ell)$ of the set of true ground states within each
sector of given orbital angular momentum $\ell,$ emerging from the
``pseudo spinless Salpeter~equation'' used in
Ref.~\cite{Hamzavi_IJMPE} (Eq.~(4) of Ref.~\cite{Hamzavi_IJMPE}),
localized by unbiased solution of Eq.~(14) of
Ref.~\cite{Hamzavi_IJMPE}.}\label{Tab:AGS}
\begin{center}\begin{tabular}{rc}\hline\hline\\[-1.5ex]
\multicolumn{1}{c}{$\ell$}&$B_{\rm p}(n_r=0,\ell)
\left[\mbox{fm}^{-1}\right]$\\[1.5ex]\hline\\[-1.5ex]
0&$-2.91894$\\1&$-0.31469$\\2&$-0.13084$\\3&$-0.06893$\\
4&$-0.04068$\\[1.5ex]\hline\hline\end{tabular}$\quad$
\begin{tabular}{rc}\hline\hline\\[-1.5ex]
\multicolumn{1}{c}{$\ell$}&$B_{\rm p}(n_r=0,\ell)
\left[\mbox{fm}^{-1}\right]$\\[1.5ex]\hline\\[-1.5ex]
5&$-0.02552$\\6&$-0.01653$\\7&$-0.01083$\\8&$-0.00706$\\
9&$-0.00451$\\[1.5ex]\hline\hline\end{tabular}$\quad$
\begin{tabular}{rc}\hline\hline\\[-1.5ex]
\multicolumn{1}{c}{$\ell$}&$B_{\rm p}(n_r=0,\ell)
\left[\mbox{fm}^{-1}\right]$\\[1.5ex]\hline\\[-1.5ex]
10&$-0.00275$\\11&$-0.00156$\\12&$-0.00078$\\13&$-0.00030$\\
14&$-0.00006$\\[1.5ex]\hline\hline\end{tabular}\end{center}
\end{table}

\item Assuming the numerical results reproduced in Tables 1 and 2
of Ref.~\cite{Hamzavi_IJMPE} to be quoted in units of
$\mbox{fm}^{-1},$ for nonvanishing orbital angular momentum
quantum number $\ell>0$ some energy levels of
Ref.~\cite{Hamzavi_IJMPE} violate more or less severely the limits
given in Table~\ref{Tab:RYP}: recall that increasing the dimension
$d$ of one's trial space may lower~and thus improve variational
upper limits, so any such discrepancy is doomed to become~more
virulent.\item Moreover, in Ref.~\cite{Hamzavi_IJMPE} a
regrettably not successful attempt was made to recover, from the
approximate analytic solution to the pseudo spinless Salpeter
equation subsumed by Eqs.~(14) and (19) of
Ref.~\cite{Hamzavi_IJMPE}, the associated solution for the
nonrelativistic limit, by wiping out in these expressions all
traces of that notorious $O(\bm{p}^4)$ free-energy~term:
\begin{itemize}\item For a Yukawa coupling constant being equal
to unity, $y=1,$ Eq.~(14) of Ref.~\cite{Hamzavi_IJMPE} simplifies
to a linear relation from which the binding energies $B_{\rm p}$
can be read~off:$$B_{\rm
p}(n_r,\ell)=-\frac{\left[m-2\left(n_r+\ell+1\right)^2b\right]^2}
{4\left(n_r+\ell+1\right)^2m} \qquad (y=1)\ .$$\item For Yukawa
coupling constants different from unity ($y\ne1$), Eq.~(14) of
Ref.~\cite{Hamzavi_IJMPE} yields an equation quadratic in the
energy $B_{\rm p},$ with two roots easily worked out. The
existence of \emph{real\/} solutions for $B_{\rm p}$ at all
depends on the involved parameters.\end{itemize}The asserted
result given in Eq.~(21) of Ref.~\cite{Hamzavi_IJMPE} matches,
however, none of the above. Clearly, all such findings have to be
confronted with the numerical solution \cite{Lucha98} of the
Schr\"odinger equation with Yukawa potential listed in the fourth
column of Table~\ref{Tab:RYP}.\item The involvement of the Yukawa
range parameter $b$ in a change of variables required in the
course of derivation of the set of approximate solutions offered
in Ref.~\cite{Hamzavi_IJMPE}~renders these solutions for
eigenvalues and eigenfunctions rather doubtful for several
reasons:\begin{itemize}\item The nonrelativistic binding energies
$B_{\rm p}$ arising from Eq.~(14) of Ref.~\cite{Hamzavi_IJMPE} for
the case of arbitrary $y\ne1$ exhibit a peculiar dependence
$B_{\rm p}(b)$ on this parameter $b$: \emph{Both\/} of these
eigenvalue solutions vanish in the (Coulomb-type) limit $b\to0$
and decrease, for large values of $b$ (in the limit $b\to\infty$),
like $B_{\rm p}(b)\propto b^2;$~thus, at least for sufficiently
large $b,$ they both predict $B_{\rm p}(b_2)<B_{\rm p}(b_1)$ for
$b_1<b_2,$~in contrast to theory-guided intuition, and they
violate the Coulomb lower limit~of~Sec.~\ref{Eq:LEB}.\item The
limit $b\to0$ reproduces the nonrelativistic Coulomb levels merely
for $y=1.$\end{itemize}\end{itemize}By the above observations, we
are led to conclude that the bulk of approximations imposed on the
spinless Salpeter equation in order to arrive at approximate
semianalytical solutions --- encoded in an implicit relation
providing the bound-state energies (Eq.~(14) of
Ref.~\cite{Hamzavi_IJMPE}) and an explicit expression for the
associated eigenfunctions (Eq.~(19) of Ref.~\cite{Hamzavi_IJMPE})
--- lead to unsatisfactory, or poor, characterizations of the
spectrum of the spinless Salpeter~equation.

\section{Summary and Concluding Remarks}With due satisfaction, we
realize that we have at our disposal a variety of sophisticated
and highly efficient techniques that enable us to draw a
sufficiently clear picture of the solutions to be expected for the
spinless Salpeter equation. With these rigorous boundary
conditions at hand, we are able to subject a proposed approximate
solution to a detailed~scrutiny~with respect to its trustability.
It is a pity that not all findings for the Yukawa case pass
this~test.


\small\begin{thebibliography}{30}
\bibitem{BSEa}H.~A.~Bethe and E.~E.~Salpeter, Phys.~Rev.~{\bf 82}
(1951) 309.
\bibitem{BSEb}M.~Gell-Mann and F.~Low,~Phys.~Rev.~{\bf 84} (1951)
350.
\bibitem{BSEc}E.~E.~Salpeter and H.~A.~Bethe, Phys.~Rev.~{\bf 84}
(1951) 1232.
\bibitem{Lucha00-HO}R.~L.~Hall, W.~Lucha, and F.~F.~Sch\"oberl,
J.~Phys.~A {\bf 34} (2001) 5059, arXiv:hep-th/0012127.
\bibitem{Lucha01-DMAIa}R.~L.~Hall, W.~Lucha, and F.~F.~Sch\"oberl,
J.~Math.~Phys.~{\bf 42} (2001) 5228, arXiv:hep-th/0101223.
\bibitem{Lucha01-DMAIb}R.~L.~Hall, W.~Lucha, and F.~F.~Sch\"oberl,
Int.~J.~Mod.~Phys.~A {\bf 17} (2002) 1931, arXiv:hep-th/0110220.
\bibitem{Lucha02-sum}R.~L.~Hall, W.~Lucha, and F.~F.~Sch\"oberl,
J.~Math.~Phys.~{\bf 43} (2002) 5913, arXiv:math-ph/0208042.
\bibitem{Lucha02-DMAII}R.~L.~Hall, W.~Lucha, and F.~F.~Sch\"oberl,
Int.~J.~Mod.~Phys.~A {\bf 18} (2003) 2657, arXiv:hep-th/0210149.
\bibitem{Martin88}A.~Martin, Phys.~Lett.~B {\bf 214} (1988) 561.
\bibitem{MMT1}M.~Reed and B.~Simon, \emph{Methods of Modern
Mathematical Physics IV: Analysis of Operators\/} (Academic Press,
New York, 1978).
\bibitem{MMT2}A.~Weinstein and W.~Stenger, \emph{Methods of
Intermediate Problems for Eigenvalues --- Theory and
Ramifications\/} (Academic Press, New York/London, 1972).
\bibitem{MMT3}W.~Thirring, \emph{A Course in Mathematical Physics
3: Quantum Mechanics of Atoms and Molecules\/} (Springer, New
York/Wien, 1990).
\bibitem{Lucha:Q&Aa}W.~Lucha and F.~F.~Sch\"oberl, Phys.~Rev.~A {\bf
60} (1999) 5091, arXiv:hep-ph/9904391.
\bibitem{Lucha:Q&Ab}W.~Lucha and F.~F.~Sch\"oberl,
Int.~J.~Mod.~Phys.~A {\bf 15} (2000) 3221, arXiv:hep-ph/9909451.
\bibitem{Lucha92:MEP}W.~Lucha, H.~Rupprecht, and F.~F.~Sch\"oberl,
Phys.~Rev.~D {\bf 45} (1992) 1233.
\bibitem{Lucha94:Como}W.~Lucha and F.~F.~Sch\"oberl, in
\emph{Proceedings of the International Conference on Quark
Confinement and the Hadron Spectrum\/}, edited by N.~Brambilla and
G.~M.~Prosperi (World Scientific, River Edge, NJ, 1995) p.~100,
arXiv:hep-ph/9410221.
\bibitem{Lucha96:CCC}W.~Lucha and F.~F.~Sch\"oberl, Phys.~Lett.~B
{\bf 387} (1996) 573, arXiv:hep-ph/9607249.
\bibitem{Lucha97:LagBd}W.~Lucha and F.~F.~Sch\"oberl,
Phys.~Rev.~A {\bf 56} (1997) 139, arXiv:hep-ph/9609322.
\bibitem{Lucha99:talk1}W.~Lucha and F.~F.~Sch\"oberl,
Int.~J.~Mod.~Phys.~A {\bf 14} (1999) 2309, arXiv:hep-ph/9812368.
\bibitem{Lucha99:talk2}W.~Lucha and F.~F.~Sch\"oberl, Fizika B {\bf
8} (1999) 193, arXiv:hep-ph/9812526.
\bibitem{Lucha04:TWR}W.~Lucha and F.~F.~Sch\"oberl, Recent
Res.~Dev.~Phys.~{\bf 5} (2004) 1423, arXiv:hep-ph/0408184.
\bibitem{Lucha14:SRWSP}W.~Lucha and F.~F.~Sch\"oberl,
Int.~J.~Mod.~Phys.~A {\bf 29} (2014) 1450057,
arXiv:1401.5970~[hep-ph].
\bibitem{Lucha14:QCD@Work}W.~Lucha and F.~F.~Sch\"oberl,
arXiv:1407.4624 [hep-ph].
\bibitem{Lucha14:SRHP}W.~Lucha and F.~F.~Sch\"oberl,
arXiv:1408.4957 [hep-ph], Int.~J.~Mod.~Phys.~A (in press).
\bibitem{AS}M.~Abramowitz and I.~A.~Stegun (eds.), \emph{Handbook
of Mathematical Functions\/} (Dover, New York, 1964).
\bibitem{Bateman}Bateman Manuscript Project, A.~Erd\'elyi et
al., \emph{Higher Transcendental Functions\/} (McGraw--Hill, New
York, 1953), Vol.~II.
\bibitem{Herbst}I.~W.~Herbst, Commun.~Math.~Phys.~{\bf 53} (1977)
285; {\bf 55} (1977) 316 (addendum).
\bibitem{Martin89}A.~Martin and S.~M.~Roy, Phys.~Lett.~B {\bf 233}
(1989) 407.
\bibitem{Bargmann}V.~Bargmann, Proc.~Natl.~Acad.~Sci.~USA {\bf
38} (1952) 961.
\bibitem{Hamzavi_IJMPE}M.~Hamzavi, S.~M.~Ikhdair, and M.~Solaimani,
Int.~J.~Mod.~Phys.~E {\bf 21} (2012) 1250016, arXiv:1203.1747
[quant-ph] (various mistakes in the printed publication have been
corrected, later on, in arXiv:1203.1747v2 [quant-ph]).
\bibitem{Hassanabadi_CPB}S.~Hassanabadi, M.~Ghominejad,
S.~Zarrinkamar, and H.~Hassanabadi, Chin.~Phys.~B {\bf 22} (2013)
060303.
\bibitem{Lucha98}W.~Lucha and F.~F.~Sch\"oberl,
Int.~J.~Mod.~Phys.~C {\bf 10} (1999) 607, arXiv:hep-ph/9811453.
\end{thebibliography}
\end{document}